\documentclass[journal]{IEEEtran}

\usepackage{amsmath,amssymb,verbatim,cite,amsopn,graphicx,citesort,url,color}

\newcommand{\hhat}{\hat{h}}
\newcommand{\htilde}{\tilde{h}}

\newcommand{\sn}{\sigma^2_n}
\newcommand{\se}{\sigma^2_e}

\newcommand{\SNR}{\text{SNR}_{\text{eff}}}

\newcommand{\Clb}{C_{\rm{LB}}}

\newcommand{\AuthorOne}{Xiangyun Zhou}

\newcommand{\ThankTwo}{X. Zhou is with Research School of Engineering, The Australian National
University (Email: xiangyun.zhou@anu.edu.au). This work was supported under
Australian Research Council's Discovery Projects funding scheme (project number
DP140101133).}

\newtheorem{Proposition}{Proposition}
\newtheorem{Corollary}{Corollary}

\begin{document}

\title{Training-Based SWIPT: \\
Optimal Power Splitting at the Receiver}

\author{
\authorblockN{\AuthorOne
\thanks{\ThankTwo} }}

\maketitle

\begin{abstract}
We consider a point-to-point system with simultaneous wireless information and power transfer (SWIPT) over a block fading channel. Each transmission block consists of a training phase and a data transmission phase. Pilot symbols are transmitted during the training phase for channel estimation at the receiver. To enable SWIPT, the receiver adopts a power-splitting design, such that a portion of the received signal is used for channel estimation or data detection, while the remaining is used for energy harvesting. We optimally design the power-splitting ratios for both training and data phases to achieve the best ergodic capacity performance while maintaining a required energy harvesting rate. Our result shows how a power-splitting receiver can make the best use of the received pilot and data signals to obtain the optimal SWIPT performance.

\end{abstract}

\begin{keywords}

Simultaneous wireless information and power transfer, power splitting, training, channel estimation.

\end{keywords}

\section{Introduction}

Recently, the concept of simultaneous wireless information and power transfer (SWIPT) has drawn considerable attention~\cite{zhang_13twc,zhou_13tcom,liu_13tcom,popovski_13tcom,xiang_12wcl,huang_13tsp,ng_13twc,krikidis_12cl,nasir_13twc,ding_13spl,krikidis_14tcom}. With simple circuit designs, the receiver is able not only to decode the information carried by the RF signal but also to harvest energy from the same signal. Two practical receiver designs that have been widely accepted are \textit{time switching} and \textit{power splitting}~\cite{zhang_13twc}. With the time-switching design, e.g., in~\cite{zhou_13tcom,krikidis_12cl,nasir_13twc}, the receiver is either in the information decoding mode or the energy harvesting mode at any point in time. For this to happen, new frame structures must be designed to include energy harvesting time slots. On the other hand, the power-splitting design, e.g., in~\cite{zhou_13tcom,liu_13tcom,huang_13tsp,ng_13twc,nasir_13twc,ding_13spl,krikidis_14tcom}, enables the receiver to split the received signal into two streams, one going to the information decoding circuit and the other going to the energy harvesting circuit. When splitting the signal, the power of the signal is also divided. The basic power-splitting design requires no change in the conventional communication systems apart from the receiver circuit.

Current studies on SWIPT systems often assume perfect channel knowledge with a few exceptions considering imperfect channel knowledge at the transmitter, e.g.~\cite{ng_14twc}. On the other hand, imperfect channel estimation at the receiver has not yet been considered. For communications over time-varying fading channels, pilot symbols are periodically transmitted to facilitate channel estimation at the receiver and the estimation is never perfect in practice. The tradeoff in resource allocation between channel training and data transmission has been investigated extensively for conventional communication systems with information transfer only~\cite{hassibi_03tit,pohl_05twc,zhou_09tsp}. Some recent independent research has also studied the resource allocation between training and energy transfer in multi-antenna systems without considering information transfer~\cite{yang_arxiv,zeng_arxiv}. In a SWIPT system, both information and energy transfers are required, hence, how to achieve the best tradeoff in resource allocation between channel estimation, data detection and energy harvesting becomes an interesting open problem. In this work, we study such a tradeoff by focusing on the power-splitting design at the receiver.

We consider a training-based SWIPT system. Each transmission block starts with a training phase followed by a data transmission phase. Considering a block fading channel, the system aims to achieve the best ergodic capacity performance whilst maintaining a target energy harvesting rate. To this end, we optimally design the power-splitting ratios during both training and data phases, denoted by $\rho_p$ and $\rho_d$. Specifically, $\rho_p$ controls the resource allocation between channel estimation and energy harvesting during training phase, and $\rho_d$ controls the resource allocation between data detection and energy harvesting during data phase. We propose both non-adaptive and adaptive power-splitting designs. In the non-adaptive design, $\rho_p$ and $\rho_d$ have fixed values for all blocks. In the adaptive design, $\rho_p$ is fixed while $\rho_d$ is dynamically chosen according to the estimated channel gain in each block. The main contributions of this work are summarized as follows:
\begin{itemize}
  \item One novel aspect of this work is the consideration of power splitting during the training phase. Our result shows that, when the training resource is limited, the receiver should use most, if not all, power for channel estimation during the training phase, and leave the burden on energy harvesting to the data phase. The optimal values of $\rho_p$ and $\rho_d$ are generally very different, which implies the importance of having different power-splitting designs for training and data phases.
  \item The adaptive power-splitting design results in a significantly improved capacity performance, as compared to the non-adaptive design, when the required energy harvesting rate is moderate to large. For the adaptive design, we also analytically compare the optimal values of $\rho_d$ with perfect and imperfect channel estimation and find them to be fundamentally different. For example, one should use all power for energy harvesting when the estimated channel gain is sufficiently small, while, in the case of perfect channel estimation, one should use all power for data detection when the channel gain is sufficiently small.
\end{itemize}

\section{System Model} \label{sec:}

We consider a single-antenna point-to-point system where the receiver makes use of the RF signal sent from the transmitter to obtain information and harvest energy. The wireless channel experiences block-wise Rayleigh fading. Each transmission block starts with a training phase having $L_p$ pilot symbols, followed by a data phase having $L_d$ data symbols. Note that both $L_p$ and $L_d$ are integers not smaller than 1. The transmit power is fixed and denoted as $P$. The channel gain is assumed to remain constant during one block and change to an independent value in the next block.

In order to receive information and harvest energy simultaneously, the receiver employs a power-splitting architecture~\cite{zhang_13twc}: upon receiving a signal $y$, it splits the signal into two streams with a power ratio $\rho$, hence, $\sqrt{\rho}y$ is used for baseband processing (i.e., channel estimation or data detection) and $\sqrt{1-\rho}y$ is used for energy harvesting. We denote $\rho_p\in[0,1]$ and $\rho_d\in[0,1]$ as the power-splitting parameters for pilot signals and data signals, respectively. Note that the receiver noise comes from both the antenna noise at the RF band and the down conversion plus baseband noise. In practice, the antenna noise is usually much smaller than the down conversion plus baseband noise~\cite{liu_13tcom}, hence the antenna noise is ignored in this work for simplicity.

\subsection{Channel Estimation}

During the training phase, the received symbol used for channel estimation is given by\vspace{-1mm}
\begin{eqnarray}\label{eq:}
        y_p = \sqrt{\rho_p} h x_p + n,
\end{eqnarray}
where $h$ is the channel gain modelled as a zero-mean complex Gaussian random variable with unit variance, $x_p$ is the transmitted pilot symbol, and $n$ is the receiver noise modelled as a zero-mean complex Gaussian random variable with variance $\sn$. With the variance of the channel gain normalized to unity, the value of $P$ represents the combined effect of the actual transmit power and path loss.

We consider the MMSE estimator~\cite{kay_93} for channel estimation based on the $L_p$ received symbols. Denote the channel estimate and the estimation error as $\hhat$ and $\htilde$, respectively, with $h = \hhat + \htilde$. Both $\hhat$ and $\htilde$ are zero-mean complex Gaussian random variables. The variance of $\htilde$ is given by~\cite{hassibi_03tit}
\begin{eqnarray}\label{eq:}
        \se = \frac{\sn}{\sn+\rho_p P L_p}.
\end{eqnarray}
From the orthogonality property of the MMSE estimator, the variance of $\hhat$ is given as $1-\se$.

\subsection{Data Transmission}

During the data phase, the received symbol used for data detection is given by\vspace{-1mm}
\begin{eqnarray}\label{eq:}
        y_d &=& \sqrt{\rho_d} h x_d + n, \\
        &=& \sqrt{\rho_d} \hhat x_d + \sqrt{\rho_d} \htilde x_d + n,
\end{eqnarray}
where $x_d$ is the transmitted data symbol.

In this work, we use the ergodic capacity to measure the performance of data transmission, which is an appropriate metric for delay tolerant applications. The exact ergodic capacity expression under imperfect channel estimation is unknown. To overcome this problem, a popular approach is to use an accurate lower bound on the ergodic capacity as the performance measure instead. We adopt a widely-used ergodic capacity lower bound~\cite{hassibi_03tit}, given as
%\footnote{Note that a pre-log factor of $L_d/(L_p+L_d)$ needs to be included in $\Clb$ if one wants to take the training overhead into account when measuring throughput (which does not affect the results in this paper).}
\begin{eqnarray}\label{eq:}
        \Clb &=& \mathbb{E} \left\{\log \Big(1 + \frac{\rho_d P |\hhat|^2 }{\sn+\rho_d P \se}\Big) \right\},
\end{eqnarray}
where $\mathbb{E}\{\cdot\}$ is the expectation operator. Since the above lower bound was shown to be an accurate approximation of the exact ergodic capacity~\cite{zhou_09tsp}, we will simply refer to $\Clb$ as the ergodic capacity. Note that a pre-log factor of $L_d/(L_p+L_d)$ can be included in $\Clb$ to take the training overhead into account when measuring the throughput (which does not affect the results in this paper).

\subsection{Energy Harvesting}

The total amount of energy harvested during one block is given by $\eta \big((1-\rho_p)P L_p |h|^2 + (1-\rho_d)P L_d |h|^2\big)$, where $\eta$ is the energy conversion efficiency. By averaging over the realizations of the channel gain, the average amount of power harvested (per symbol) is given by
\begin{eqnarray}\label{eq:PowHar}
        Q &=& \eta \frac{\mathbb{E}\{(1-\rho_p)P L_p |h|^2 + (1-\rho_d)P L_d |h|^2 \}}{L_p+L_d},
\end{eqnarray}

\subsection{Problem Formulation}

We focus on the design at the receiver only and consider the following optimization problem: how to optimally split the power of the received signal during both training and data phases so that the ergodic capacity is maximized while maintaining an acceptable energy harvesting rate, i.e.,\vspace{-1mm}
\begin{subequations}\label{eq:Problem}
\begin{eqnarray}
        \max_{\rho_p,\rho_d}& \,\,\,\,\,\,& \Clb \\
        \text{s.t.} && Q \geq Q_0\\
        && 0\leq \rho_p \leq 1\\
        && 0\leq \rho_d \leq 1
\end{eqnarray}
\end{subequations}
where $Q_0$ is the minimum required power to be harvested on average. For convenience and without loss of generality, we assume $\eta = 1$. This is equivalent to making $Q_0$ represent the required power to be directed into the energy harvesting circuit before any conversion loss. The feasible range of $Q_0$ is given by $Q_0\in[0,P]$, since the maximum power that can be harvested is $P$. Therefore, we will assume $Q_0\in[0,P]$ when solving the above optimization problem. Another observation one can make is that the energy harvesting inequality constraint can be simplified to an equality constraint, i.e., $Q = Q_0$, since instead of harvesting extra power beyond $Q_0$, one can use such power for channel estimation or data detection so that the ergodic capacity is further increased.

\section{Non-Adaptive Power Splitting Design}

In this section, we consider a non-adaptive design where the power-splitting parameters, $\rho_p$ and $\rho_d$, are fixed for all blocks. Let us rewrite the ergodic capacity expression as
\begin{eqnarray}\label{eq:}
        \Clb &=& \mathbb{E} \left\{\log \Big(1 + \frac{\rho_d P (1-\se)}{\sn +\rho_d P \se}|h_0|^2 \Big) \right\},
\end{eqnarray}
where $h_0$ is a random variable having the same distribution as $\hhat$ but with unit variance instead of $1-\se$. Let us further define an effective SNR as
\begin{eqnarray}
        \SNR(\rho_p,\rho_d) &=& \frac{\rho_d P (1-\se)}{\sn+\rho_d P \se}\\
        &=& \frac{\rho_d P \Big(1-\frac{\sn}{\sn+\rho_p P L_p}\Big)}{1+\rho_d P \frac{\sn}{\sn+\rho_p P L_p}}.\label{eq:SNRb}
\end{eqnarray}
It is clear that the non-adaptive power-splitting parameters affect the ergodic capacity only through $\SNR$. Therefore, the optimization problem in (\ref{eq:Problem}) reduces to
\begin{subequations}
\begin{eqnarray}\label{eq:ProblemP1}
       \text{P1}: \,\,\,\,\,\,\max_{\rho_p,\rho_d}& \,\,\,\,\,\,& \SNR(\rho_p,\rho_d)\\
        \text{s.t.} && Q = Q_0\\
        && 0\leq \rho_p \leq 1\\
        && 0\leq \rho_d \leq 1
\end{eqnarray}
\end{subequations}

\begin{Proposition}\label{Proposition:1}
{\em{The optimal non-adaptive power splitting design that solves P1 is given by\vspace{-1mm}
 \begin{eqnarray}
\rho^*_p &=& \left\{ \begin{array}{ll}
 \rho_{p,lb}, &\mbox{ \!if  $\rho_{p,r}<\rho_{p,lb}$,} \\
  \rho_{p,r}, &\mbox{ \!if  $\rho_{p,lb} \leq \rho_{p,r} \leq \rho_{p,ub}$,}\\
  \rho_{p,ub}, &\mbox{ \!if  $\rho_{p,r}>\rho_{p,ub}$,}
       \end{array} \right.\label{eq:rhopOpt}\\
       \rho^*_d &=& 1 - \frac{Q_0(L_p+L_d)}{P L_d} + (1-\rho^*_p)\frac{L_p}{L_d},
       \end{eqnarray}
where
\begin{eqnarray}
        \rho_{p,lb}  &=& \max \left\{ 0\, ,\, 1- \frac{Q_0(L_p+L_d)}{P L_p} \right\}, \label{eq:rhoLB}\\
        \rho_{p,ub} &=&  \min \left\{ 1\, ,\, 1- \frac{Q_0(L_p+L_d)}{P L_p} + \frac{L_d}{L_p}\right\},\label{eq:rhoUB}\\
        \rho_{p,r} &=& \left\{ \begin{array}{ll}
 \frac{L_p+1}{2 L_p}\Big(1-\frac{Q_0}{P}\Big), &\mbox{ \!if  $L_d = 1$,} \\
  \frac{L_d+\kappa-\sqrt{L_d(\kappa+L_d)(\kappa+1)}}{L_p (1-L_d)P/\sn}, &\mbox{ \!if  $L_d > 1$,}
       \end{array} \right.\label{eq:rhopr}
\end{eqnarray}
and
\begin{eqnarray}\label{eq:kappa}
        \kappa  &=& \frac{P-Q_0}{\sn}(L_d+L_p).
\end{eqnarray}
}}
\end{Proposition}

{\em{Proof}}: See Appendix~A.

To obtain further insights into the optimal power splitting solution, we consider the scenario where the transmit power is sufficiently large and the energy harvesting constraint also scales linearly, i.e., $P/\sn \rightarrow \infty$ and $Q_0/P = c$ where $c\in(0,1)$ is an arbitrary constant.

\begin{Corollary}\label{Corollary:1}
{\em{With $P/\sn \rightarrow \infty$ and $Q_0/P = c$, the solution to the optimal non-adaptive power splitting design is given in Proposition 1 with\vspace{-1mm}
\begin{eqnarray}\label{eq:}
        \rho_{p,r} &=& \frac{L_p+L_d}{L_p (1+\sqrt{L_d})} (1-c).
\end{eqnarray}
}}
\end{Corollary}

From Corollary 1, we see that $\rho_{p,r}$ decreases as $L_p$ increases. This implies that the optimal design is to use less power for channel estimation but more power for energy harvesting when the training length increases. This observation agrees with intuition: Firstly, it is important to achieve a reasonably accurate channel estimation which will benefit data detection significantly. When the training resource is limited (i.e., small $L_p$), it is wise to use most, if not all, resource for channel estimation. On the other hand, when the channel estimation is already accurate by using a portion of the resource, allocating additional power to further improve channel estimation gives marginal improvement in data detection, and hence, it is better to use the additional power for energy harvesting instead.

\section{Adaptive Power Splitting Design}

In the previous section, the power-splitting parameters are designed to have constant values in every block regardless of the channel quality in each block. In fact, the power splitting during the training phase has to be fixed since the channel quality is unknown to the receiver prior to channel estimation. On the other hand, the power splitting during data phase can be designed adaptively according to the channel estimate. In this section, we consider such an adaptive design with the value of $\rho_d$ optimally chosen in each block while $\rho_p$ remains fixed for all blocks.

Firstly, since $\hhat$ and $\htilde$ are uncorrelated~\cite{hassibi_03tit}, the average amount of harvested power defined in (\ref{eq:PowHar}) can be simplified to (with the assumption of $\eta = 1$)
\begin{eqnarray}\label{eq:}
        \!\!\!Q \!\!\!&=&\!\!\! \frac{ (1\!-\!\rho_p)P L_p + \mathbb{E} \Big\{ (1\!-\!\rho_d)P L_d (|\hhat|^2\!+\!\se) \Big\}}{L_p+L_d},
\end{eqnarray}
where the expectation is taken over the realizations of $|\hhat|^2$ and the power-splitting parameter $\rho_d$ is a function of $|\hhat|^2$.

To solve the optimization problem in (\ref{eq:Problem}), we first optimize the adaptive parameter $\rho_d$ for a given value of $\rho_p$, and then find the optimal value of $\rho_p$. The two-step problem can be expressed as
\begin{subequations}
\begin{eqnarray}
       \text{P2.1}: \,\,\,\max_{\rho_d} & & \Clb = \mathbb{E} \left\{\log \Big(1+ \frac{\rho_d P |\hhat|^2 }{\sn+\rho_d P \se}\Big) \right\}\nonumber\\
       \\
        \text{s.t.} && \mathbb{E} \Big\{ (1-\rho_d)(|\hhat|^2+\se) \Big\} = \xi \label{eq:EHconstD}\\
        && 0\leq \rho_d \leq 1,
\end{eqnarray}
\end{subequations}
where $\rho_d$ is a function of $|\hhat|^2$ and $\xi$ is related to $\rho_p\in[\rho_{p,lb},\rho_{p,ub}]$ by $\xi = \frac{Q_0(L_p+L_d)-(1-\rho_p)P L_p}{P L_d}$.
\begin{subequations}
\begin{eqnarray}\label{eq:}
       \text{P2.2}: \,\,\,\,\,\,\max_{\rho_p}& & \Clb = \mathbb{E} \left\{\log \Big(1+ \frac{\rho^*_d P |\hhat|^2 }{\sn+\rho^*_d P \se}\Big) \right\}\nonumber\\
       \\
        \text{s.t.} && \rho_{p,lb} \leq \rho_p \leq \rho_{p,ub},
\end{eqnarray}
\end{subequations}
where $\rho^*_d$ denotes the optimal adaptive power-splitting policy during data phase, i.e., the solution to Problem P2.1. The imposed range of $\rho_p\in[\rho_{p,lb},\rho_{p,ub}]$ with $\rho_{p,lb}$ and $\rho_{p,ub}$ given in (\ref{eq:rhoLB}) and (\ref{eq:rhoUB}), can be understood as follows: For any feasible value of $\rho_p$ smaller than $\rho_{p,lb}$ (if exists), the average power harvested during the training phase alone is already larger than $Q_0$, which is not optimal. For any feasible value of $\rho_p$ larger than $\rho_{p,ub}$ (if exists), the required average power to be harvested during the data phase cannot be achieved even with $\rho_d = 0$, which makes the optimization problem P2.1 infeasible. Hence, the optimal value of $\rho_p$ must satisfy $\rho_p\in[\rho_{p,lb},\rho_{p,ub}]$. Consequently, we have $\xi\in[0,1]$.

\subsection{Solution to Problem P2.1: Adaptive Data Power Splitting with Imperfect Channel Estimation}

We first solve Problem P2.1, which has its own practical meaning: what is the optimal adaptive power splitting policy during data transmission that maximizes the ergodic capacity while satisfying an energy harvesting constraint during data transmission. The optimal policy in the special case of perfect channel estimation, i.e., $\se=0$, was studied in~\cite{liu_13tcom}. In what follows, we solve for the optimal policy with imperfect channel estimation, i.e., $\se\in(0,1)$.

Firstly, we discuss two trivial cases: When $\xi = 0$, no energy needs to be harvested, hence the optimal policy is to use all power for data detection, i.e., $\rho^*_d = 1\, \forall \, |\hhat|^2$. When $\xi = 1$, all available energy must be harvested, hence the only feasible policy is $\rho^*_d = 0\, \forall \, |\hhat|^2$. Next, we present the solution for $\xi\in(0,1)$.

\begin{Proposition}\label{Proposition:2}
{\em{The optimal adaptive data power splitting design that solves P2.1 with $\xi\in(0,1)$ is given by
  \begin{eqnarray}\label{eq:rhodOpt}
\rho^*_d &=& \left\{ \begin{array}{ll}
0, &\mbox{ \!if  $\rho_{d,r}<0$,} \\
  \rho_{d,r}, &\mbox{ \!if  $0 \leq \rho_{d,r} \leq 1$,} \\
  1, &\mbox{ \!if  $\rho_{d,r}>1$,}
       \end{array} \right.
\end{eqnarray}
where
\begin{eqnarray}\label{eq:rhodr}
        \!\!\!\rho_{d,r} \!\!\!&=&\!\!\! \frac{-(\nu\!+\!\se)\!+\!\sqrt{(\nu\!+\!\se)^2-4\se\nu
        (1\!-\!\frac{P|\hhat|^2}{\sn\nu\lambda^*})}}{2\se \nu P/\sn}
\end{eqnarray}
and
\begin{eqnarray}\label{eq:}
        \nu = (|\hhat|^2+\se).
\end{eqnarray}
The constant $\lambda^*\in(0,P/\sn)$ in (\ref{eq:rhodr}) can be found via a simple bisection method to satisfy the energy harvesting constraint in (\ref{eq:EHconstD}).}}
\end{Proposition}

{\em{Proof}}: See Appendix~B.

Generally, the value of $\rho^*_d$ is not a monotonic function of $|\hhat|^2$. This can be seen by studying the limiting case of $|\hhat|^2\rightarrow0$ and $|\hhat|^2\rightarrow \infty$ as stated in the following corollary.

\begin{Corollary}\label{Corollary:2}
{\em{Still assume a nontrivial energy harvesting constraint with $\xi\in(0,1)$. When $|\hhat|^2\rightarrow0$, we have $\rho^*_d = 0$. When $|\hhat|^2\rightarrow\infty$, we have $\rho^*_d \rightarrow 0$. Hence, the value of $\rho^*_d$ is not a monotonic function of $|\hhat|^2$.
}}
\end{Corollary}

In the case of perfect channel estimation, as studied in~\cite{liu_13tcom}, the solution to the optimal data power splitting policy is given by
 \begin{eqnarray}\label{eq:OptPerfect}
\rho^*_d &=& \left\{ \begin{array}{ll}
\frac{1}{|h|^2}\Big(\frac{1}{\lambda^*}-\frac{\sn}{P}\Big), &\mbox{ \!if  $|h|^2\geq \frac{1}{\lambda^*}-\frac{\sn}{P}$,} \\
 1, &\mbox{ \!if  $|h|^2 < \frac{1}{\lambda^*}-\frac{\sn}{P}$,}
       \end{array} \right.
\end{eqnarray}
with the constant $\lambda^*\in(0,P/\sn)$ found via a bisection method to satisfy the energy harvesting constraint. From (\ref{eq:OptPerfect}), we see that with perfect channel estimation, $\rho^*_d = 1$ when the channel gain is sufficiently small. This is in contrast to the result with imperfect channel estimation which says that $\rho^*_d = 0$ when the estimated channel gain is sufficiently small. In addition, the value of $\rho^*_d$ is monotonically decreasing as the channel gain gets larger in the case of perfect channel estimation, while such a monotonic relation does not generally exist in the case of imperfect channel estimation.

\subsection{Solution to Problem P2.2}

After obtaining the optimal solution to adaptive data power-splitting policy given in the previous subsection, we can now numerically solve the optimal training power-splitting policy as stated in Problem P2.2. Specifically, for any given value of $\rho_p$, the corresponding ergodic capacity is computed by solving Problem P2.1. Then, the optimal value of $\rho_p\in[\rho_{p,lb},\rho_{p,ub}]$ that maximizes the ergodic capacity can be found via a one-dimensional line search.

\section{Numerical Results}

In this section, we present numerical results to illustrate the optimal power-splitting policies and the optimal ergodic capacity performance. The following parameter settings are used in all plots: $P = 100$, $\sn = 1$, $L_p + L_d = 100$. The ergodic capacity is obtained from 50000 simulation runs in Matlab. In addition, we have chosen base-$e$ logarithm to present the results, hence, the capacity unit is nats per channel use. The energy harvesting constraint is expressed in terms of $Q_0/P$, i.e., the required average power to be harvested normalized by the transmit power.

\begin{figure}[!t]
\centering
\includegraphics[width=1\columnwidth]{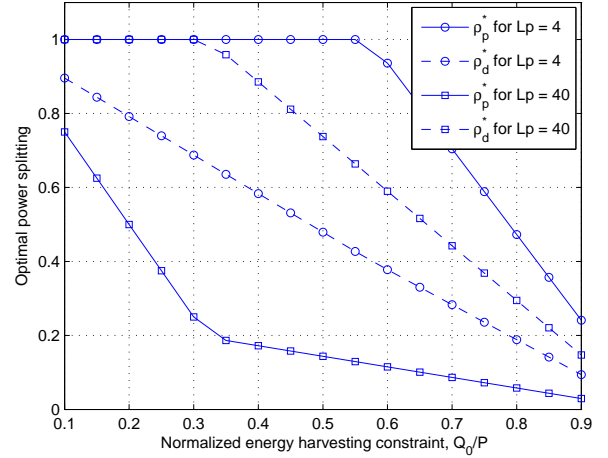}
\vspace{-3mm} \caption{The optimal non-adaptive power splitting policies during both training and data phases.}
\label{fig:opt_splitting_nonadaptive}
\end{figure}

We first look at the non-adaptive power-splitting design. Fig.~\ref{fig:opt_splitting_nonadaptive} shows the optimal power splitting during both the training phase and data phase, i.e., $\rho^*_p$ and $\rho^*_d$, for two different training lengths. When the training length is small, it is important to make a good use of the received pilot symbols to achieve accurate channel estimation. In other words, the receiver should use a significant portion of the received power for channel estimation, and leave most, if not all, of the burden on energy harvesting to the data phase. As we see in Fig.~\ref{fig:opt_splitting_nonadaptive}, when $L_p = 4$, all pilot power is used for channel estimation when $Q_0/P < 0.55$. On the other hand, when the training length is large, the receiver only needs a portion of pilot power to achieve accurate channel estimation. In this case, more energy can be harvested during training, reducing the burden on energy harvesting during data phase. As shown in Fig.~\ref{fig:opt_splitting_nonadaptive}, when $L_p = 40$, a larger fraction of power is used for energy harvesting during the training phase as compared to the data phase.

\begin{figure}[!t]
\centering
\includegraphics[width=1\columnwidth]{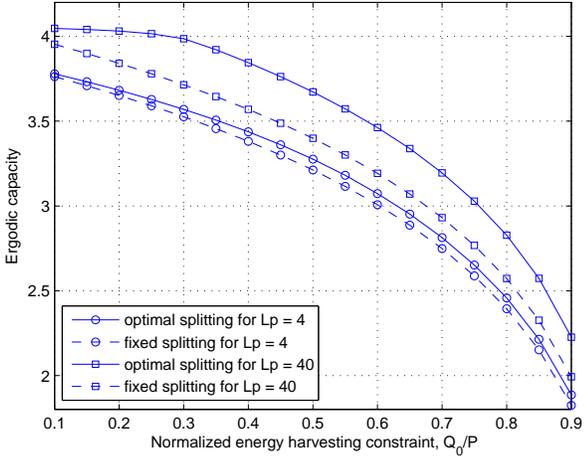}
\vspace{-3mm} \caption{The ergodic capacity achieved with non-adaptive power splitting.}
\label{fig:capacity_nonadaptive}
\end{figure}

Fig.~\ref{fig:opt_splitting_nonadaptive} also shows that the optimal power-splitting policy is generally quite different from a simple fixed power-splitting policy with the same power-splitting ratio used for all symbols (i.e., $\rho_p = \rho_d = 1 - Q_0/P$). In Fig.~\ref{fig:capacity_nonadaptive}, we compare the ergodic capacity performance between the optimal and fixed power-splitting policies. Generally, the capacity difference is notable, especially when the training length is large. This confirms that it is important to differentiate the training and data phases when designing the power-splitting policy for good SWIPT performance.

\begin{figure}[!t]
\centering
\includegraphics[width=1\columnwidth]{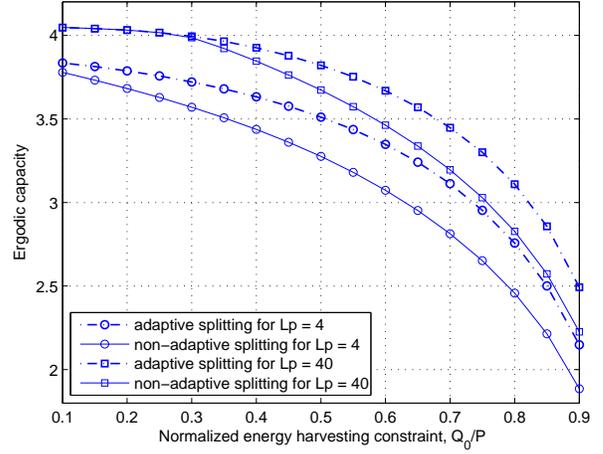}
\vspace{-3mm} \caption{The ergodic capacity achieved with both non-adaptive and adaptive power splitting.}
\label{fig:capacity_comparison}
\end{figure}

Next, we compare the capacity performance between non-adaptive and adaptive power-splitting designs in Fig.~\ref{fig:capacity_comparison}. Again, we see some notable capacity difference, especially when the energy harvesting constraint is moderate to large. When the energy harvesting constraint is small, the scenario is similar to the one without any energy harvesting constraint, and hence, it is optimal to use almost all power for channel estimation and data detection. The adaptive power-splitting design only makes sense when the energy harvesting constraint is non-negligible.

\section{Conclusion and Future Work}

In this work, we focused on the design of a power-splitting receiver in a SWIPT system with channel estimation errors. The optimal power-splitting policies during the training and data phases were derived to maximize the ergodic capacity whilst meeting an energy harvesting constraint. Compared to a simple fixed power-splitting policy that does not differentiate training and data phases, the optimal policies can achieve a notable capacity improvement even with just the non-adaptive design. If the adaptive power-splitting design is adopted, further improvement on the ergodic capacity is also significant when the required energy harvesting rate is moderate to large.

This work only considered a baseline model: the point-to-point link. Future work can extend the study to multiuser system with one-to-many SWIPT. The results on optimal power splitting in this work can be directly applied to time division multiple access (TDMA) systems. However, the extension becomes non-trivial with multi-antenna transmission, especially when quantized channel information feedback and multiuser precoding are considered. Another interesting extension is to further optimize the training length and compare the result with the optimal training length in systems with either information transfer only or energy transfer only.

\appendix

\section{Appendix}

\subsection{Proof of Proposition 1}
\label{app:PP1}

Since both $\rho_p$ and $\rho_d$ are constants and $\mathbb{E}\{|h|^2\}=1$, we use the energy harvesting equality constraint to write $\rho_d$ in terms of $\rho_p$ as
\begin{eqnarray}\label{eq:rhoPandD}
        \rho_d &=& 1 - \frac{Q_0(L_p+L_d)}{P L_d} + (1-\rho_p)\frac{L_p}{L_d}.
\end{eqnarray}
The optimal values of $\rho_p$ and $\rho_d$ must satisfy the above equality. Since $0\leq \rho_d \leq 1$, using (\ref{eq:rhoPandD}) we obtain the feasible range of $\rho_p$ as $\rho_{p,lb} \leq \rho_p \leq \rho_{p,ub}$. Substituting (\ref{eq:rhoPandD}) into $\SNR$ in (\ref{eq:SNRb}), the optimization problem over a single parameter $\rho_p$ can be written as $\max_{\rho_p\in[\rho_{p,lb},\rho_{p,ub}]}\,\SNR(\rho_p)$. Taking the derivative of $\SNR(\rho_p)$ w.r.t. $\rho_p$ and solve for the roots, we obtain
\begin{eqnarray}
       \rho_{p,r} &=& \left\{ \begin{array}{ll}
 \frac{L_p+1}{2 L_p}\Big(1-\frac{Q_0}{P}\Big), &\mbox{ \!if  $L_d = 1$,}\\
  \frac{L_d+\kappa \,\pm\, \sqrt{L_d(\kappa+L_d)(\kappa+1)}}{L_p (1-L_d)P/\sn}, &\mbox{ \!if  $L_d > 1$.}
       \end{array} \right.\nonumber
\end{eqnarray}
Specifically, there is a single (positive) root when $L_d = 1$, while there are two roots when $L_d > 1$. Since $\kappa \geq 0$, it is not difficult to show that the only non-negative root when $L_d > 1$ is the one with the `$-$' sign in front of $\sqrt{\cdot}$, which must correspond to a maximum point. By taking the feasible range of $\rho_p$ into account, the optimal value of $\rho_p$ is obtained as in (\ref{eq:rhopOpt}).

\subsection{Proof of Proposition 2}
\label{app:PP2}

Problem P2.1 is a convex optimization problem in $\rho_d(|\hhat|^2)$. Here, we provide a proof similar to the one in~\cite{liu_13tcom}. First, we write the Lagrangian as
\begin{eqnarray}
        L(\rho_d,\lambda) &=& \mathbb{E} \left\{\log \Big(1+ \frac{\rho_d P |\hhat|^2 }{\sn+\rho_d P \se}\Big) \right\} \nonumber\\
        &&+ \lambda\left(\mathbb{E} \Big\{ (1-\rho_d)(|\hhat|^2+\se) \Big\}-\xi \right),\nonumber
\end{eqnarray}
where $\lambda$ is the Lagrange multiplier. Assuming $\lambda\neq0$. This is because $\rho^*_d = 1 \,\,\forall \,|\hhat|^2$, with $\lambda=0$, which cannot satisfy the energy harvesting constraint with $\xi\in(0,1)$. Consider the Lagrange dual function: $\max_{\rho_d\in[0,1]} \,\,\, L(\rho_d,\lambda)$, which can be decoupled into parallel subproblems, each for a realization of $|\hhat|^2$. For a given $|\hhat|^2$, the corresponding subproblem reduces to
\begin{eqnarray}\label{eq:subproblem}
        \max_{\rho_d\in[0,1]}\,\,\, \log \Big(1+ \frac{\rho_d P |\hhat|^2 }{\sn+\rho_d P \se}\Big)
        +\lambda (1-\rho_d)(|\hhat|^2+\se).
\end{eqnarray}
Taking the derivative of (\ref{eq:subproblem}) w.r.t. $\rho_d$, we obtain the following quadratic equation
\begin{eqnarray}\label{eq:quadratic}
              \frac{P\se\nu}{\sn}\rho^2_d + (\nu+\se)\rho_d + \frac{\sn}{P}-\frac{|\hhat|^2}{\nu\lambda} = 0.
\end{eqnarray}
Note that we have assumed imperfect channel estimation, i.e., $\se>0$. If there is no channel estimation error, i.e., $\se=0$, the second order term disappears and there is a single root to the above equation.

Before solving for the roots, we investigate the feasible range of $\lambda$ as follows: Since $\nu > |\hhat|^2 \geq 0$, there is no real positive root to (\ref{eq:quadratic}) if $\lambda < 0$ or $\lambda \geq P/\sn$, for all realizations of $|\hhat|^2$. When there is no positive root, the optimal value of $\rho_d\in[0,1]$ that maximizes the objective function is always 0, which cannot satisfy the energy harvesting equality constraint with $\xi\in(0,1)$. Therefore, we conclude that the feasible range of $\lambda$ is $(0,P/\sn)$, which is assumed in the remainder of the proof.

The roots to (\ref{eq:quadratic}) are given by
\begin{eqnarray}
               \rho_{d,r} &=& \frac{-(\nu+\se) \,\pm\,\sqrt{(\nu+\se)^2-4\se\nu
        (1-\frac{P|\hhat|^2}{\sn\nu\lambda^*})}}{2\se \nu P/\sn},\nonumber
\end{eqnarray}
With $\lambda > 0$, it is easy to show that the expression inside $\sqrt{\cdot}$ is positive, hence, both roots are real. Furthermore, it is easy to see that the only root that can be positive is the one with the `$+$' sign in front of $\sqrt{\cdot}$. Therefore, the optimal value of $\rho_d\in[0,1]$ is expressed in (\ref{eq:rhodOpt}) for any given $\lambda$.

The remaining step is to find $\lambda$ that satisfies the energy harvesting constraint. This can be done iteratively by solving for $\rho^*_d$ with a given $\lambda$ and check against the energy harvesting equality constraint in (\ref{eq:EHconstD}), then updating $\lambda$ via a bisection method until the equality constraint is met.

\bibliographystyle{IEEEtran}

\end{document}